\documentclass[showpacs,showkeys,twoside,eqsecnum,twocolumn]{revtex4-1}
\usepackage{amsfonts,amsmath,amssymb,bm,hyperref,graphicx}

\begin{document}


\title{Axially and spherically symmetric solitons in warm plasma}

\author{Maxim Dvornikov}
\email{maxdvo@izmiran.ru}
\affiliation{N.~V.~Pushkov Institute of Terrestrial Magnetism,
Ionosphere and Radiowave Propagation, \\ 142190, Troitsk, Moscow
Region, Russia}

\date{\today}

\begin{abstract}
We study the existence of stable axially and spherically symmetric
plasma structures on the basis of the new nonlinear
Schr\"{o}dinger equation (NLSE) accounting for nonlocal electron
nonlinearities. The numerical solutions of NLSE having the form of
spatial solitions are obtained and their stability is analyzed. We
discuss the possible application of the obtained results to the
theoretical description of natural plasmoids in the atmosphere.
\end{abstract}



\maketitle

\section{Introduction}

The studies of stable spatial solitons is an important problem of
contemporary physics~\cite{InfRow90}. There are numerous
manifestations of stable solitonic solutions of nonlinear
equations in two and three spatial dimensions in nonlinear
optics~\cite{Kro04}, solid state~\cite{Sku06} and plasma
physics~\cite{ShuEli10}.

In plasma physics it was established~\cite{Zak72} that the
nonlinear electron-ion interaction leads to the modulation
instability and results in the collapse of a Langmiur
wave~\cite{Zak72,Gol84}. In contrast to the electron-ion
interaction, the nonlinear electron-electron interactions, studied
in Refs.~\cite{Kuz76,SkoHaa80}, were shown to stabilize the
evolution of a Langmuir wave packet, making possible the existence
of stable spatial plasma structures.

It is convenient to describe the evolution of nonlinear Langmuir
wave packets on the basis of a nonlinear Schr\"{o}dinger equation
(NLSE)~\cite{SulSul99}. The nonlinear terms studied in
Refs.~\cite{Kuz76,SkoHaa80} are local since their contributions to
NLSE contain only a certain power of the electric field
amplitude. The nonlocal terms in NLSE, derived in
Refs.~\cite{Lit75,DavYakZal05}, are also important when, e.g., a
wave packet is very steep. Under certain conditions these nonlocal
electron nonlinearities can arrest the Lamgmuir collapse. It is
suggested in Ref.~\cite{DavYakZal05} that a nonlocal NLSE is a
theoretical model for stable spatial plasma structures obtained in
a laboratory~\cite{CheWon85}.

Besides the nonlocal electron nonlinearities taken into account in
Ref.~\cite{DavYakZal05} there are analogous contributions to NLSE
originating from the electron pressure term. These terms have the
same order of magnitude as those considered in
Ref.~\cite{DavYakZal05} and are important for plasma with nonzero
electron temperature. Note that nonlinear waves in warm plasma
were also studied in Ref.~\cite{InfRow87}. In the present work we
carefully study these additional nonlinearities and examine their
contribution to NLSE.

This paper is organized as follows. In Sec.~\ref{DYN}, on the
basis of the system of nonlinear plasma equations, we examine
electrostatic plasma waves having axial and spherical symmetry.
Then, in Sec.~\ref{LOCNON}, we briefly review the previous studies
of the influence of electron nonlinearities on the dynamics of
Langmuir waves in plasma. The new NLSE, describing stable spatial
plasma structures, taking into account the nonlocal nonlinear
terms is derived in Sec.~\ref{NONLOCNON}. In Sec.~\ref{NUM} we
analyze solutions of this NLSE numerically. We consider the
possible application of our results to the theoretical explanation
of the existence of atmospheric and ionospheric plasmoids in
Sec.~\ref{APPL}. Finally we summarize our results in
Sec.~\ref{CONCL}.

The description of plasma waves in frames of Lagrange variables is
presented in Appendix~\ref{LAGRANGE}.

\section{The dynamics of spatial Langmuir solitons\label{DYN}}

In this section we study axially and spherically symmetric waves
in plasma accounting for local and nonlocal electron
nonlinearities. We derive a new NLSE
which is shown to have solitonic solutions.

To describe electrostatic waves in isotropic warm plasma, in which
the magnetic field is equal to zero, $\mathbf{B} = 0$, we start
from the system of nonlinear hydrodynamic equations,
\begin{gather}
  \frac{\partial n_e}{\partial t} + \nabla \cdot (n_e \mathbf{v}_e) = 0,
  \notag
  \\
  \frac{\partial v_{ej}}{\partial t} +
  (\mathbf{v}_e \cdot \nabla) v_{ej} =
  -\frac{e}{m} E_j - \frac{1}{m n_e} \nabla_i p_{ij}
  \notag
  \\
  \frac{\partial \mathbf{E}}{\partial t} =
  4 \pi e (n_e \mathbf{v}_e - n_i \mathbf{v}_i),
  \notag
  \\
  (\nabla \cdot \mathbf{E}) = - 4 \pi e (n_e - n_i),
  \label{hydrodyneq}
\end{gather}
where $n_{e,i}$ are the densities of electrons and ions,
$\mathbf{v}_{e,i}$ are their velocities, $\mathbf{E}$ is the
amplitude of the electric field, $m$ is the electron mass, $e>0$
is the proton charge, and
\begin{equation}\label{pressure}
  p_{ij} = m \int
  (\mathbf{v} - \mathbf{v}_e)_i (\mathbf{v} - \mathbf{v}_e)_j
  f_e \mathrm{d}^3 \mathbf{v},
\end{equation}
is the pressure tensor which is calculated using the electron
distribution function $f_e$. Note that Eq.~\eqref{hydrodyneq}
follows from more general Vlasov kinetic equation for the function
$f_e$.

It is known that Eq.~\eqref{hydrodyneq} allows small amplitude
Langmuir oscillations on the frequency $\omega_p = \sqrt{4 \pi n_0
e^2/m}$, where $n_0$ is the unperturbed electron density. However,
since Eq.~\eqref{hydrodyneq} is nonlinear, the higher harmonics
generation is possible. Therefore we can look for the solution of
Eq.~\eqref{hydrodyneq} in the following form:
\begin{align}\label{s12def}
  n_e = & n_0 + n_s + n_f^{(1)} + n_f^{(2)} + \dotsb,
  \notag
  \\
  \mathbf{E} = & \mathbf{E}_s + \mathbf{E}_f^{(1)} + \mathbf{E}_f^{(2)} + \dotsb,
  \notag
  \\
  \mathbf{v}_e = & \mathbf{v}_s + \mathbf{v}_f^{(1)} + \mathbf{v}_f^{(2)} +\dotsb,
\end{align}
where we separate different time scales,
%
\begin{align}\label{12def}
  n_f^{(1)} = & n_1 e^{- \mathrm{i} \omega_p t} +
  n_1^{*{}} e^{\mathrm{i} \omega_p t}, 
  \notag
  \\
  \mathbf{v}_f^{(1)} = &
  \mathbf{v}_1 e^{- \mathrm{i} \omega_p t} +
  \mathbf{v}_1^{*{}} e^{\mathrm{i} \omega_p t}, 
  \notag
  \\
  \mathbf{E}_f^{(1)} = &
  \mathbf{E}_1 e^{- \mathrm{i} \omega_p t} +
  \mathbf{E}_1^{*{}} e^{\mathrm{i} \omega_p t}, 
  \notag
  \\
  n_f^{(2)} = & n_2 e^{- 2 \mathrm{i} \omega_p t} +
  n_2^{*{}} e^{2 \mathrm{i} \omega_p t}
  \notag
  \\
  \mathbf{v}_f^{(2)} = &
  \mathbf{v}_2 e^{- 2 \mathrm{i} \omega_p t} +
  \mathbf{v}_2^{*{}} e^{2 \mathrm{i} \omega_p t},
  \notag
  \\
  \mathbf{E}_f^{(2)} = &
  \mathbf{E}_2 e^{- 2 \mathrm{i} \omega_p t} +
  \mathbf{E}_2^{*{}} e^{2 \mathrm{i} \omega_p t}.
\end{align}
%
The functions $n_s$, $\mathbf{E}_s$, and $\mathbf{v}_s$ in
Eq.~\eqref{s12def} and the amplitude functions $n_{1,2}$,
$\mathbf{v}_{1,2}$ and $\mathbf{E}_{1,2}$ in Eq.~\eqref{12def} are
supposed to vary slowly on the $1/\omega_p$ time scale. Moreover
we suggest that, e.g., $n_0 \gg n_1 \gg n_{s,2}$ etc, i.e. slowly
varying functions, marked with index ``$s$", and the amplitudes of
the second harmonic are much smaller than the corresponding
amplitudes of the main oscillation.

We neglect the velocity of ions in Eq.~\eqref{hydrodyneq} and
suggest that ion density is represented as $n_i = n_0 + n$, where
the perturbation $n$ is also a slowly varying function on the
$1/\omega_p$ time scale. Note that $n$ does not necessarily
coincide with $n_s$.

We will study electrostatic plasma oscillations in two or three
dimensions. Thus we assume radially symmetric quantities in
Eq.~\eqref{hydrodyneq},
\begin{equation}\label{rdependence}
  \begin{pmatrix}
    \mathbf{v}_e \\
    \mathbf{E} \
  \end{pmatrix} =
  \mathbf{e}_r
  \times
  \begin{pmatrix}
    v_e(r,t) \\
    E(r,t) \
  \end{pmatrix},
  \quad
  \begin{pmatrix}
    n_e \\
    n_i \\
  \end{pmatrix} =
  \begin{pmatrix}
    n_e(r,t) \\
    n_i(r,t) \\
  \end{pmatrix},
\end{equation}
where $r$ is the radial coordinate and $\mathbf{e}_r$ is the basis
vector in spherical or cylindrical coordinate system.

\subsection{Local electron nonlinearities\label{LOCNON}}

The contribution of electron nonlinearities to the evolution of
Langmiur waves in frames of the
model~\eqref{hydrodyneq}-\eqref{rdependence} was taken into
account in Refs.~\cite{Kuz76,SkoHaa80} and the following equation
for the description of the main oscillation amplitude
$\mathbf{E}_1$ was obtained:
\begin{align}\label{NLSE0}
  \mathrm{i} \dot{\mathbf{E}}_1 + &
  \frac{3}{2} \omega_p r_\mathrm{D}^2 \nabla (\nabla \cdot \mathbf{E}_1) -
  \frac{\omega_p}{2n_0} n \mathbf{E}_1
  \notag
  \\
  & -
  \frac{\beta(d)}{12 \pi m n_0 \omega_p} \frac{\mathbf{E}_1|\mathbf{E}_1|^2}{r^2}
  = 0,
\end{align}
where $r_\mathrm{D} = \sqrt{T_e/4\pi e^2 n_0}$ is the Debye
length, $T_e$ is the electron temperature, and $\beta(d) =
(d-1)(4-d)/2$, $d$ is the dimension of space.

Eq.~\eqref{NLSE0} should be supplied with the wave equation for
the ion motion~\cite{Che87},
\begin{equation}\label{ions}
  \left(
    \frac{\partial^2}{\partial t^2} - c_s^2 \Delta
  \right) n =
  \frac{\Delta |\mathbf{E}_1|^2}{4 \pi M},
\end{equation}
where $c_s = \sqrt{(T_e + \gamma_i T_i)/M}$ is the sound velocity,
$T_i$ is the ions temperature, $\gamma_i$ is the heat capacity
ratio for ions, $M$ is the ion mass, and $\Delta$ is the Laplace
operator.

At the absence of electron nonlinearities [the last term in
Eq.~\eqref{NLSE0}] the system~\eqref{NLSE0} and~\eqref{ions}
corresponds to the Zakharov equations~\cite{Zak72}. It should be
noted that the contribution of local electron nonlinearities is
washed out from Eq.~\eqref{NLSE0} in one dimensional case $d=1$.

The Zakharov equations are known to reveal the collapse of a
Langmuir wave packet~\cite{Gol84}: the size of a wave packet is
contracting and the amplitude of the electric field is growing. It
was shown in Refs.~\cite{Kuz76,SkoHaa80} that Langmuir collapse
can be arrested and stable spatial plasma structures can appear in
two and three dimensions, $d=2,3$, since the second nonlinear term
in Eq.~\eqref{NLSE0} is defocusing.

\subsection{Nonlocal electron nonlinearities\label{NONLOCNON}}

The influence of electron nonlinearities on the dynamics of a
Langmuir collapse was further studied in Ref.~\cite{DavYakZal05}.
Using the relation between slowly varying electron density $n_s$
and the perturbation of ion density $n$,
\begin{equation}
  n_s =
  n + \frac{\Delta |\mathbf{E}_1|^2}{4 \pi e^2} +
  r_\mathrm{D}^2 \Delta n_s,
\end{equation}
which was obtained in Ref.~\cite{SkoHaa80}, we can approximately
find $n_s$ as
\begin{equation}\label{nsn}
  n_s \approx
  n + \frac{\Delta |\mathbf{E}_1|^2}{4 \pi e^2} +
  r_\mathrm{D}^2 \Delta n.
\end{equation}
Note that the last term in Eq.~\eqref{nsn}, $\sim r_\mathrm{D}^2
\Delta n$, which is important at rapidly varying ion density, was
omitted in Refs.~\cite{Kuz76,SkoHaa80}.

Using Eqs.~\eqref{ions}, \eqref{nsn} and supposing that $T_i \ll
T_e$, one obtains the new nonlinear term in the left hand side of
Eq.~\eqref{NLSE0} (see Ref.~\cite{DavYakZal05}),
\begin{equation}\label{Dav}
  \frac{\mathbf{E}_1 \Delta |\mathbf{E}_1|^2}
  {32 \pi n_0 m \omega_p}.
\end{equation}
This new contribution was shown in Ref.~\cite{DavYakZal05} to
arrest the Langmuir collapse. Moreover the nonlocal
nonlinearity~\eqref{Dav} is more effective in preventing the
collapse compared to that found in Refs.~\cite{Kuz76,SkoHaa80}. It
should be also noted that the new nonlinear term predicted in
Ref.~\cite{DavYakZal05} does not disappear in one dimensional
case. The nonlinearity analogous to that in Eq.~\eqref{Dav}, $\sim
\mathbf{E}_1 \Delta |\mathbf{E}_1|^2$, appears in the Zakharov
equations with quantum effects~\cite{HaaShu09}, while taking into
account the quantum Bohm potential.

The nonlocal electron nonlinearities, analogous to that studied in
Ref.~\cite{DavYakZal05}, can follow not only from Eq.~\eqref{nsn}.
We can consider the contribution of the slowly varying electron
density to the electron pressure~\eqref{pressure}. Analogously to
Eqs.~\eqref{s12def} and~\eqref{12def} one can discuss the
decomposition of the distribution function $f_e$ proposed in
Ref.~\cite{Kuz76},
\begin{align}\label{feexp}
  f_e = & f_0 + f_s +
  f_1 e^{- \mathrm{i} \omega_p t} + f_1^{*{}} e^{\mathrm{i} \omega_p t}
  \notag
  \\
  & +
  f_2 e^{- 2\mathrm{i} \omega_p t} + f_2^{*{}} e^{2\mathrm{i} \omega_p t}
  + \dotsb,
\end{align}
where $f_0$ is the equilibrium distribution function. For
classical plasma it can be, e.g., a Maxwell distribution
corresponding to $T_e$. All the quantities in the expansion
series~\eqref{feexp} depend on $\mathbf{v} - \mathbf{v}_f^{(1)}$.

Using the result of Ref.~\cite{Kuz76} we represent $f_1$ as,
\begin{align}\label{f1}
  f_1 = & \frac{\mathrm{i}}{\omega_p}
  \bigg[
    \frac{\partial f_0}{\partial \mathbf{v}}
    (\mathbf{v} \cdot \nabla) \mathbf{v}_1 +
    \frac{\partial f_s}{\partial \mathbf{v}}
    (\mathbf{v} \cdot \nabla) \mathbf{v}_1
    \notag
    \\
    & -
    (\mathbf{v}_1 \cdot \nabla) f_s +
    \frac{e}{m} \mathbf{E}_s
    \frac{\partial f_1}{\partial \mathbf{v}}
  \bigg] + \dotsb,
\end{align}
where we drop terms containing $f_2$, $\mathbf{E}_2$, and higher
power of the first harmonic amplitudes. The function $f_s$ in
Eq.~\eqref{f1} was also found in Ref.~\cite{Kuz76} for the case of
isotropic equilibrium distribution,
\begin{equation}\label{fs}
  f_s = - v_\mathrm{T}^2 n_s
  \frac{1}{v} \frac{\mathrm{d} f_0}{\mathrm{d} v} + \dotsb,
\end{equation}
where $v_\mathrm{T} = \sqrt{T_e/m}$ is the thermal velocity of
electrons. The terms which do not contain the slowly varying
density $n_s$ are omitted in Eq.~\eqref{fs}.

Now we can express the nonlinear term $\nabla_i p_{ij}/n_e$ in
Eq.~\eqref{hydrodyneq} as
\begin{align}\label{nablap0}
  \frac{1}{n_e}\nabla_i p_{ij} = &
  \frac{1}{n_0}
  \bigg[
    \left(
      1-\frac{n_s}{n_0}
    \right)
    \nabla_i \int \mathrm{d}\mathbf{v} v_i v_j f_1
    \notag
    \\
    & -
    \frac{n_1}{n_0} \nabla_i \int \mathrm{d}\mathbf{v} v_i v_j f_s
  \bigg] e^{-\mathrm{i} \omega_p t}.
\end{align}
Using the Maxwell equilibrium distribution function,
\begin{equation}
  f_0 = n_0
  \left(
    \frac{m}{2 \pi T_e}
  \right)^{3/2}
  \exp
  \left(
    -\frac{m v^2}{2 T_e}
  \right),
\end{equation}
normalized on the unperturbed electron density, we can express the
gradient of pressure in Eq.~\eqref{nablap0} in the following form:
\begin{align}\label{nablap1}
  \nabla_i p_{ij} = &
  3 m v_\mathrm{T}^2 \nabla_i n_1 e^{-\mathrm{i} \omega_p t}
  \notag
  \\
  & -
  \frac{e v_\mathrm{T}^2}{\omega_p^2}
  e^{-\mathrm{i} \omega_p t}
  \big[
    \nabla_j (n_s \nabla_i E_{1i}) +
    \nabla_i (n_s \nabla_i E_{1j})
    \notag
    \\
    & +
    \nabla_i (n_s \nabla_j E_{1i}) +
    \nabla_j (E_{1i} \nabla_i n_s)
    \notag
    \\
    & -
    (\nabla_j n_s) (\nabla_i E_{1i}) -
    3 n_s \nabla_j (\nabla_i E_{1i})
  \big],
\end{align}
where we use the relations between the amplitudes of the main
harmonic,
\begin{equation}
  \mathbf{v}_1 = - \frac{\mathrm{i}e}{m\omega_p} \mathbf{E}_1
  \quad
  (\nabla \cdot \mathbf{E}_1) = - 4 \pi e n_1,
\end{equation}
established in Ref.~\cite{SkoHaa80}. The leading term in
Eq.~\eqref{nablap1} was derived in Refs.~\cite{Kuz76,SkoHaa80}.
The next-to-leading terms, proportional to the derivatives of
$n_s$, are important when one has rapidly varying in space wave
packets.

Using Eqs.~\eqref{nsn} and~\eqref{nablap1} we can obtain the
generalization of Eq.~\eqref{NLSE0} which takes into account the
nonlocal electron nonlineariries due to the interaction of a
Langmuir wave with the low-frequency perturbation of electron
density,
\begin{align}\label{NLSE1}
  \mathrm{i} \dot{\mathbf{E}} + &
  \frac{3}{2} \omega_p r_\mathrm{D}^2 \nabla (\nabla \cdot \mathbf{E}) -
  \frac{\omega_p}{2n_0} n \mathbf{E} -
  \frac{\beta(d)}{12 \pi m n_0 \omega_p} \frac{\mathbf{E}|\mathbf{E}|^2}{r^2}
  \notag
  \\
  & -
  \frac{v_\mathrm{T}^2}{2 n_0 \omega_p}
  \big[
    \mathbf{E} \Delta n -
    \nabla [n (\nabla \cdot \mathbf{E})] -
    \nabla_i (n \nabla_i \mathbf{E})
    \notag
    \\
    & -
    \nabla_i (n \nabla E_{i}) -
    \nabla [(\mathbf{E} \cdot \nabla) n] +
    (\nabla \cdot \mathbf{E}) \nabla n
    \notag
    \\
    & +
    3 n \nabla (\nabla \cdot \mathbf{E})
  \big] = 0,
\end{align}
where for simplicity we omit the index ``$1$": $\mathbf{E} \equiv
\mathbf{E}_1$. Eq.~\eqref{NLSE1} should be supplied with
Eq.~\eqref{ions} governing the evolution of ion density
perturbation $n$.

The nonlocal term $\sim \mathbf{E} \Delta n$ in Eq.~\eqref{NLSE1}
was derived in Ref.~\cite{DavYakZal05}. The remaining nonlocal
nonlinearities, which are of the same order of magnitude as the
term $\sim \mathbf{E} \Delta n$, were omitted in that work.

Let us study the evolution of the system~\eqref{ions}
and~\eqref{NLSE1} in the subsonic regime, when one can neglect the
second time derivative of the ion density in Eq.~\eqref{ions}.
Considering axially symmetric, $d=2$, or spherically symmetric,
$d=3$, cases [see Eq.~\eqref{rdependence}] and using the
dimensionless variables,
\begin{equation}
  x = \sqrt{\frac{3}{2}} \frac{r}{r_\mathrm{D}},
  \quad
  s = \frac{9}{4} \omega_p t,
  \quad
  \Phi = \frac{E}{\sqrt{18 \pi n_0 T_e}},
\end{equation}
we can represent the dynamics of the system~\eqref{ions}
and~\eqref{NLSE1} in the form of a single NLSE,
\begin{align}\label{NLSE2}
  \mathrm{i} \frac{\partial \Phi}{\partial s} =
  &
  -\frac{\partial}{\partial x}
  \left[
    \frac{1}{x^{d-1}}
    \frac{\partial}{\partial x}
    (x^{d-1} \Phi)
  \right] -
  \Phi |\Phi|^2 + \beta(d) \frac{\Phi |\Phi|^2}{x^2}
  \notag
  \\
  & -
  \frac{3}{2}
  \left[
    \Phi \frac{d-1}{x} -
    3 \frac{\partial \Phi}{\partial x}
  \right]  \frac{\partial}{\partial x} |\Phi|^2 = 0.
\end{align}
It should be noted that Eqs.~\eqref{NLSE1} or~\eqref{NLSE2} does
not conserve the number of plasmons,
\begin{equation}
  N_\phi = \int \mathrm{d} V |\mathbf{E}|^2,
  \quad
  \dot{N}_\phi \neq 0.
\end{equation}
This fact is because of the presence of the term $\sim \partial
\Phi/\partial x$ in Eq.~\eqref{NLSE2}.

To analyze the integrals of Eq.~\eqref{NLSE2} let us separate the
variables in Eq.~\eqref{NLSE2}: $\Phi = e^{-\mathrm{i}\lambda s}
\phi(x)$. Then making the following nonlinear gauge transformation
of the ``nonhermitian" Eq.~\eqref{NLSE2}:
\begin{equation}\label{gauge}
  \psi = \exp
  \left(
    - \frac{9}{4} |\phi|^2
  \right) \phi,
\end{equation}
we can cast it in the equivalent form,
\begin{align}\label{NLSE3}
  \lambda \psi =  &
  -\frac{\partial}{\partial x}
  \left[
    \frac{1}{x^{d-1}}
    \frac{\partial}{\partial x}
    (x^{d-1} \psi)
  \right] -
  \psi |\psi|^2 + \beta(d) \frac{\psi |\psi|^2}{x^2}
  \notag
  \\
  & -
  \frac{3}{2}
  \left[
    \frac{3}{2} \Delta_x |\psi|^2 +
    \frac{d-1}{x} \frac{\partial |\psi|^2}{\partial x}
  \right]\psi,
\end{align}
where $\Delta_x = \partial^2/\partial x^2 + (d-1)/x \times
\partial/\partial x$ is the radial part of the Laplace operator.
To derive Eq.~\eqref{NLSE3} we use the decomposition of the
transformed ``wave function" squared~\eqref{gauge}, $|\psi|^2 = (1
- \frac{9}{2} |\phi|^2 + \dotsb) |\phi|^2 \approx |\phi|^2$, and
keep only cubic terms in Eq.~\eqref{NLSE3}.

The modified NLSE~\eqref{NLSE3} is ``hermitian", i.e. it conserves
the number of plasmons,
\begin{equation}\label{Npsi}
  N_\psi = \Omega_d \int_0^\infty \mathrm{d}x \ x^{d-1} |\Psi|^2,
\end{equation}
where we restore the $s$ dependence, $\Psi = e^{-\mathrm{i}\lambda
s} \psi(x)$, and $\Omega_d = 2 \pi$, for $d = 2$, or $\Omega_d = 4
\pi$, for $d = 3$, is the solid angle. It should be also noticed
that the nonlocal nonlinearities does not disappear in one
dimensional case as the local ones studied in
Ref.~\cite{Kuz76,SkoHaa80}. The dynamics of Langmuir waves in warm
plasma in arbitrary dimensions is analyzed in
Appendix~\ref{LAGRANGE}, using Lagrange variables. It is shown
there that in case of nonzero electron temperature the nonlinear
terms does not disappear in higher dimensions $d>1$.

We can notice that Eq.~\eqref{NLSE3} is analogous to NLSE equation
derived in Ref.~\cite{DavYakZal05}. It also contains $\psi
\Delta_x |\psi|^2$ term, although the coefficient is different.
The main discrepancy is the presence of the term $\sim (d-1)/x
\times \psi \partial |\psi|^2/\partial x$. In Sec.~\ref{NUM} we
analyze the influence of this new contribution numerically.

\section{Numerical simulation\label{NUM}}

It is difficult to construct other conserved integrals of
Eq.~\eqref{NLSE3}, e.g., a Hamiltonian, independent of the number
of plasmons~\eqref{Npsi}. Therefore one has to analyze the
behaviour of solutions of this equation numerically.

Eq.~\eqref{NLSE3} should be supplied with the boundary conditions,
$\psi(0) = \psi(\infty) = 0$, and thus treated as a boundary
condition problem. We have found numerical solutions of this
problem using a boundary condition problem solver, incorporated in
the MATLAB~7.6 program. It requires an initial ``guess" function
which was chosen as
\begin{equation}\label{psig}
  \psi_g(x) = \sqrt{N_0} A_d x \exp
  \left(
    -\frac{x^2}{2\sigma^2}
  \right),
\end{equation}
where $\sigma$ is the ``width" of the function, $A_d =
1/(\sqrt{\pi}\sigma^2)$, for $d = 2$, and $A_d =
\sqrt{2/3}/(\pi^{3/4}\sigma^{5/2})$, for $d = 3$. The
function~\eqref{psig} is now normalized on the initial number of
plasmons $N_0$, to be compares with the actual number of plasmons
$N_\psi$ obtained from a numerical solution.

\begin{figure*}
  \centering
  \includegraphics[scale=1]{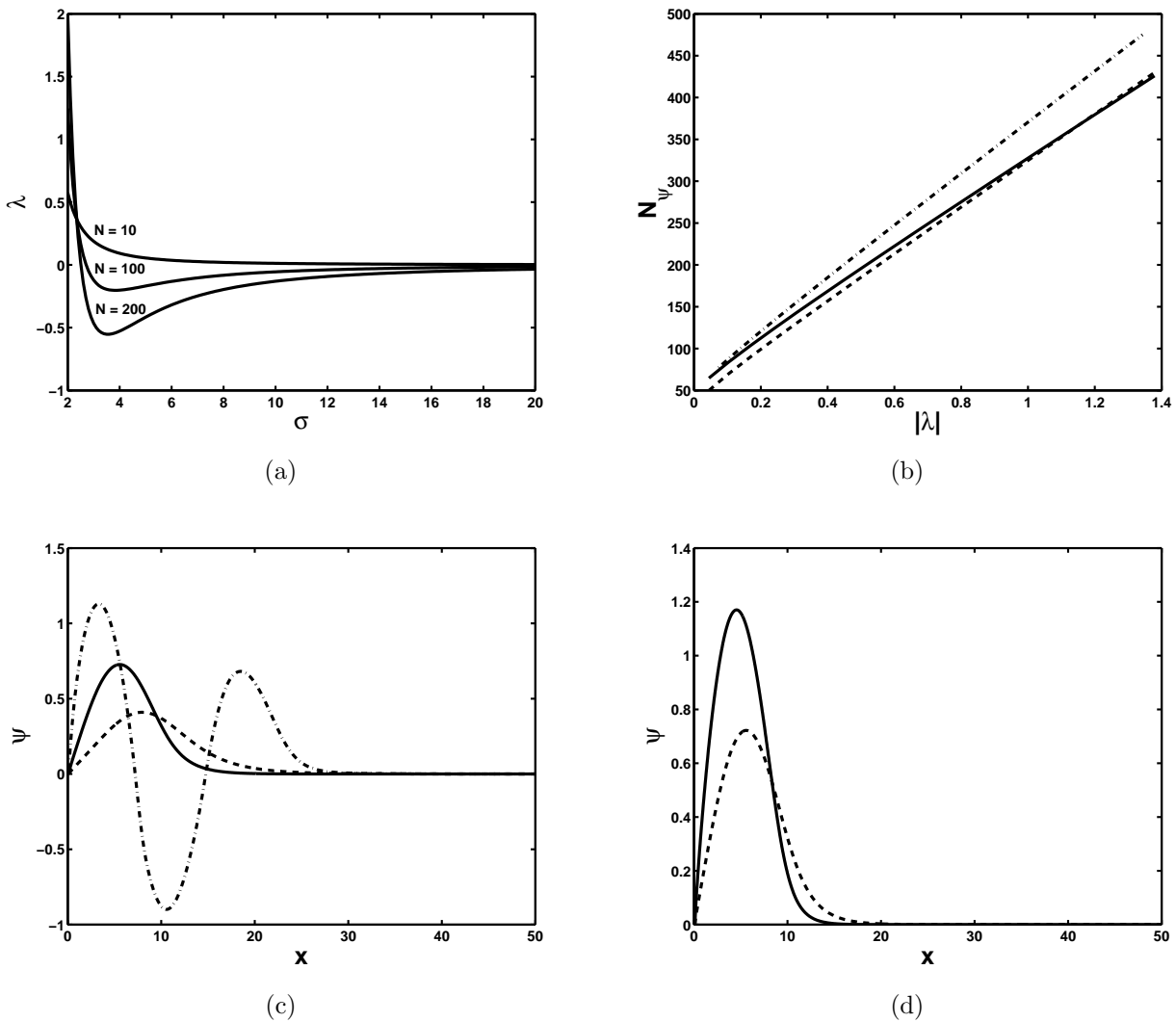}
  \caption{\label{2Dfig}
  The analysis of solutions of Eq.~\eqref{NLSE3} in 2D case.
  (a)~The function $\lambda(\sigma)$ obtained on the basis
  of Eq.~\eqref{NLSE3} using a ``guess" function
  $\psi = \psi_g$~\eqref{psig} for different values of $N_0$.
  (b)~The dependence $N_\psi(\lambda)$:
  solid line corresponds to a numerical solution of Eq.~\eqref{NLSE3},
  dashed line represents an ``analytical" curve corresponding to
  minimal $\lambda_a$, and
  dash-dotted line corresponds to a numerical solution of Eq.~\eqref{NLSE3}
  at the absence of the term
  $(d-1)/x \times \psi \partial |\psi|^2/\partial x$
  (see Ref.~\cite{DavYakZal05}).
  (c)~Examples of solutions of Eq.~\eqref{NLSE3}
  obtained for $N_0 = 100$;
  solid line: $N_\psi \approx 113$,
  $\lambda \approx - 0.20$, and $\sigma \approx 3.83$
  (optimal parameters minimizing $\lambda$),
  dashed line: $N_\psi \approx 70$,
  $\lambda \approx - 0.06$, and $\sigma \approx 9.56$, and
  dash-dotted line: $N_\psi \approx 595$,
  $\lambda \approx - 0.06$, and $\sigma \approx 2.83$.
  (d)~The same as in panel~(c) for $N_0 = 200$;
  solid line: $N_\psi \approx 210$,
  $\lambda \approx - 0.55$, and $\sigma \approx 3.55$
  (optimal parameters),
  dashed line: $N_\psi \approx 113$,
  $\lambda \approx - 0.20$, and $\sigma \approx 7.91$, and
  dash-dotted: $N_\psi \approx 113$,
  $\lambda \approx - 0.20$, and $\sigma \approx 2.64$.
  Note that in the latter case dashed and dash-dotted lines practically
  coincide.}
\end{figure*}

\begin{figure*}
  \centering
  \includegraphics[scale=1]{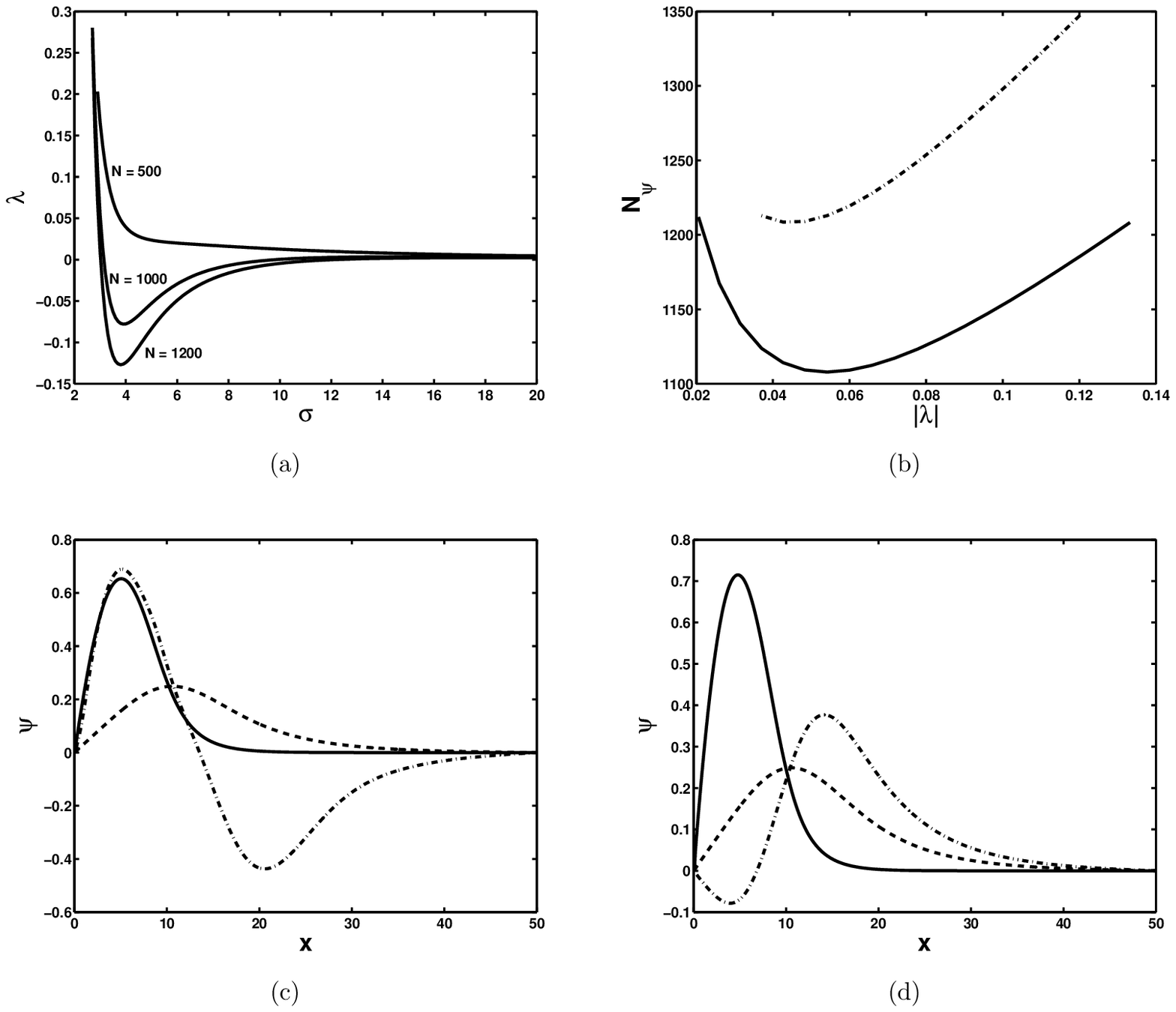}
  \caption{\label{3Dfig}
  Panels~(a) and~(b) are the same as in Fig.~\ref{2Dfig}(a,b)
  but correspond to 3D case.
  (c)~Examples of stable solutions of Eq.~\eqref{NLSE3}
  [right-handed branch in panel~(b)]
  obtained for $N_0 = 1100$;
  solid line: $N_\psi \approx 1150$,
  $\lambda \approx - 0.10$, and $\sigma \approx 3.90$
  (optimal parameters),
  dashed line: $N_\psi \approx 1429$,
  $\lambda \approx - 0.01$, and $\sigma \approx 8.22$, and
  dash-dotted line: $N_\psi \approx 1219$,
  $\lambda \approx - 0.01$, and $\sigma \approx 3.10$.
  (d)~The same as in panel~(c) for $N_0 = 1200$;
  solid line: $N_\psi \approx 1197$,
  $\lambda \approx - 0.13$, and $\sigma \approx 3.80$
  (optimal parameters),
  dashed line: $N_\psi \approx 1429$,
  $\lambda \approx - 0.01$, and $\sigma \approx 8.82$, and
  dash-dotted line: $N_\psi \approx 4354$,
  $\lambda \approx - 0.01$, and $\sigma \approx 3.04$.}
\end{figure*}

The best convergence of the numerical procedure is achieved when
$\sigma$ corresponds to the minimal value of $\lambda$ in
Eq.~\eqref{NLSE3} at the given $N_0$. This analysis is analogous
to the trial function method~\cite{And83} for minimizing of the
Hamiltonian of NLSE. In Fig.~\ref{2Dfig}(a), for $d = 2$, and in
Fig.~\ref{3Dfig}(a), for $d = 3$, we present the dependence of
$\lambda$ versus $\sigma$ for different values of $N_0$.
One can see that the function $\lambda(\sigma)$ has a minimum if
$N_0 > N_\mathrm{cr}$. In two dimensional case the critical number
of plasmons can be easily found: $N_\mathrm{cr} = 8\pi$. Note that
one can expect the stability of a soliton with respect to the
collapse if $\lambda < 0$ for $d = 2$, whereas in 3D case there is
some range of positive $\lambda$ which corresponds to uncollapsing
solutions.

The solutions which correspond to various values of $\lambda$,
$\sigma$, and $N_\psi$ are shown in Fig.~\ref{2Dfig}(c,d), for $d
= 2$, and in Fig.~\ref{3Dfig}(c,d), for $d = 3$. A ``guess"
function which does not correspond to the minimum of the function
$\lambda(\sigma)$ also gives some solitonic solution of
Eq.~\eqref{NLSE3}, however the convergence is much worse than in
the minimal $\lambda$ case. Indeed, $N_\psi$ significantly differs
from $N_0$ for such a solution. When the deviation from the
minimal $\lambda$ is big, although $\lambda$ remains to be
negative, no regular solutions can be found and the system
capsizes into chaos. Thus these ``nonoptimal"  solutions seem to
be unstable.

\begin{figure}
  \centering
  \includegraphics[scale=.4]{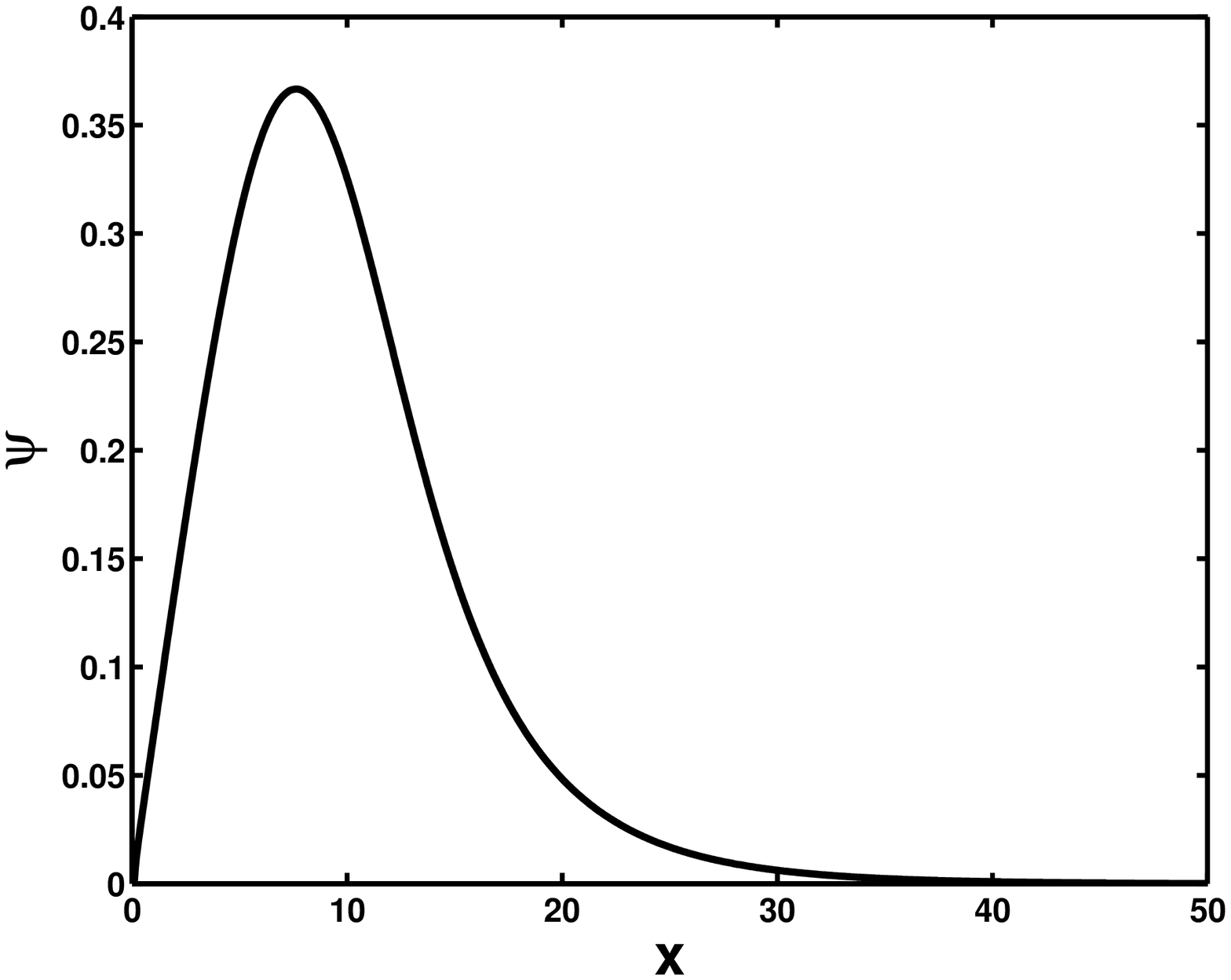}
  \caption{\label{unstable}
  An example of the unstable solution of Eq.~\eqref{NLSE3}
  [left-handed branch in Fig.~\ref{3Dfig}(b)] obtained for $N_0 = 770$.
  It corresponds to $N_\psi \approx 1174$,
  $\lambda \approx - 0.02$, and $\sigma \approx 4.20$.}
\end{figure}

To analyze the stability of the found solutions we present the
$N_\psi(\lambda)$ dependence in Fig.~\ref{2Dfig}(b), for $d = 2$,
and in Fig.~\ref{3Dfig}(b), for $d = 3$. Note these curves were
built for ``guess" functions corresponding to a minimal $\lambda$.
Applying the Vakhitov--Kolokolov criterion~\cite{SulSul99p72} to
the results shown on these plots one can conclude that the
presented solutions are stable in 2D case, whereas some unstable
solitons can exist in three dimensions. For $d = 3$, a stable
solution can be generated starting from a threshold plasmon number
$N_\mathrm{cr} \approx 1100$ and at $|\lambda| > 0.05$ (see also
the discussion in Sec.~\ref{APPL}). We show an example of a
unstable soliton in Fig.~\ref{unstable}.

In Fig.~\ref{2Dfig}(b), for $d = 2$, one can see that the new term
in Eq.~\eqref{NLSE3}, $\sim (d-1)/x \times \psi \partial
|\psi|^2/\partial x$, does not produce any significant effect
(compare solid and dash-dotted lines). In 3D case,
Fig.~\ref{3Dfig}(b), the difference is just quantitative: the
critical plasmon number and critical frequency are shifted.
Therefore our results are in agreement with
Ref.~\cite{DavYakZal05} where NLSE with a nonlocal term was
analyzed. One should also notice that the numerical curve in
Fig.~\ref{2Dfig}(b) (solid line) is in a good agreement with the
``analytical" $\lambda(N_\psi)$ dependence (dashed line),
$\lambda_a = - (N_\psi - 8 \pi)^2/(88 \pi N_\psi)$, found from
Eq.~\eqref{NLSE3} with help of the ``guess" function~\eqref{psig}.

\section{Applications\label{APPL}}

Stable spatial solitons, involving nonlocal nonlinearities,
similar to plasma structures described in the present work, were
reported to be obtained in various laboratory experiments in
plasma physics, condescended matter, and nonlinear optics (see,
e.g., reviews~\cite{Kro04,Sku06,ShuEli10} and references therein).
%
%
Another opportunity for the physical realization of plasmoids
described in Secs.~\ref{DYN} and~\ref{NUM} consists in their
implementation as a rare natural atmospheric electricity
phenomenon called a ball lightning (BL)~\cite{Ste99}.

According to the BL observations, most frequently it has a form of
a rather regular sphere with the diameter of
$(20-50)\thinspace\text{cm}$~\cite{Ste99p11}. However big BLs with
the size more than one meter were reported to
exist~\cite{BycGolNik10}. Besides spherical BL, snake-like objects
were observed~\cite{BycGolNik10}. The lifetime of BL can be up to
several minutes~\cite{Ste99p11}. The estimates of energy of BL
were obtained only in cases when it produced some destruction
while disappearing. These estimates give for the energy of BL the
value in the range (several~kJ --
several~MJ)~\cite{BycGolNik10p214}. However, since in many cases
BL disappears just fading, one should assume that a very low
energy BL can exist.

Numerous BL models are reviewed in Ref.~\cite{theoryBL}. Despite
very exotic theoretical descriptions of BL were proposed, it is
most probable that this object is a plasma based phenomenon. In
Refs.~\cite{DvoDvo,Dvo10} we developed a BL model based on radial
oscillations of electron gas in plasma, studying these
oscillations in both classical and quantum approaches. The present
work is a development of our BL model. As in our previous studies,
here we also treat BL on the basis of radial oscillations of
electrons. However, as we show in Secs.~\ref{DYN} and~\ref{NUM},
local and nonlocal electron nonlinearities as well as the
interaction of electrons with ions play an important role for the
stability of an axially or spherically symmetric plasmoid. For
example, the inhomogeneity of ion density was not accounted for in
Ref.~\cite{DvoDvo}. It should be noted that the idea that
spherically symmetric oscillations of electrons underlie the BL
phenomenon was independently put forward in Ref.~\cite{Shm03}.

\begin{figure*}
  \centering
  \includegraphics[scale=1]{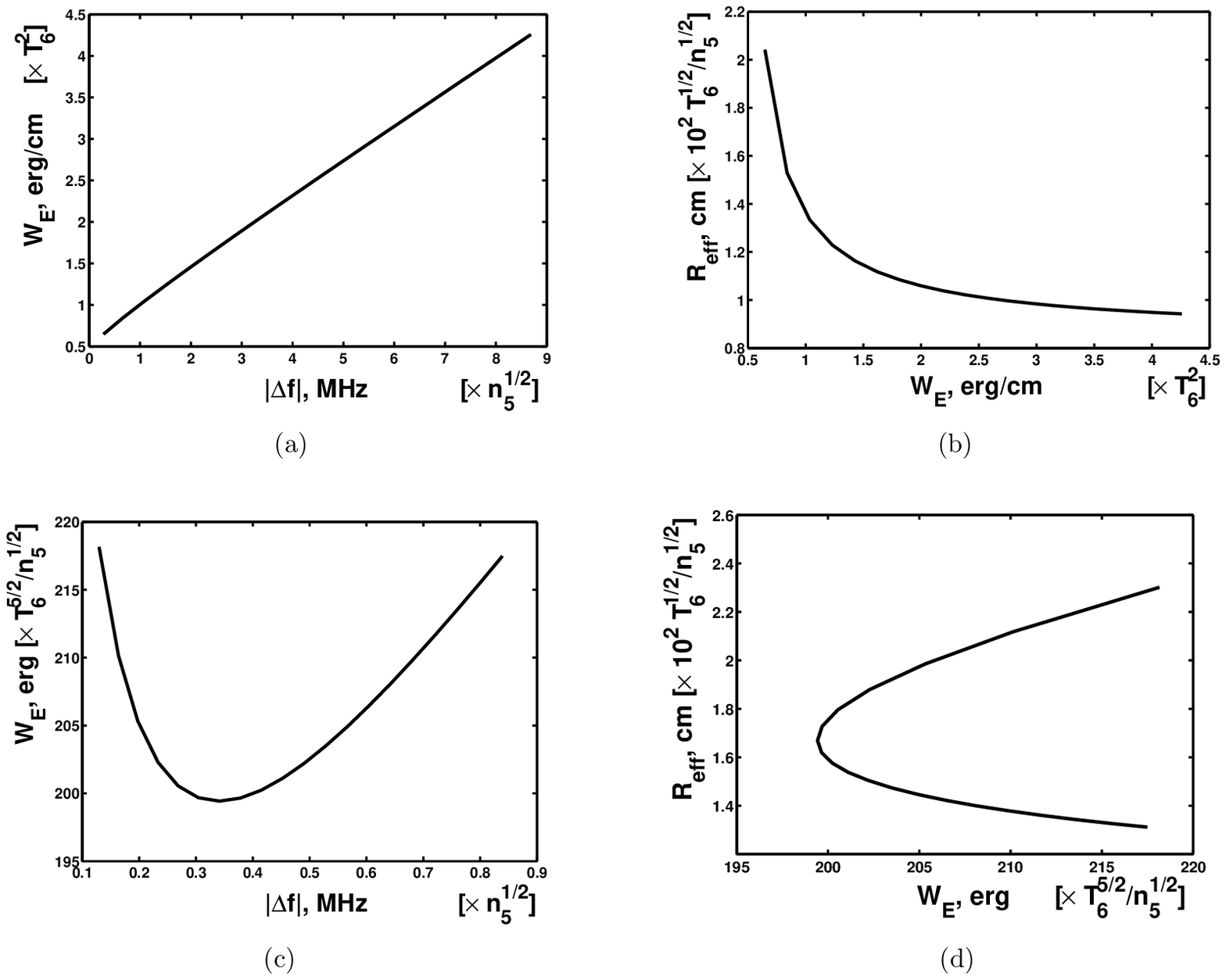}
  \caption{\label{BLprop}
  Physical characteristics of axial [panels~(a) and~(b)]
  and spherical [panels~(c) and~(d)] plasmoids expressed in terms of
  $T_6 = T_e/10^6\thinspace\text{K}$ and
  $n_5 = n_0/10^5\thinspace\text{cm}^{-3}$.
  (a)~The energy density of electric field vs. the frequency shift.
  (b)~The effective radius vs. the energy density.
  (c)~The total energy  of electric field vs. the frequency shift.
  Stable plasmoids correspond to the right-handed branch.
  (d)~The effective radius vs. the total energy.
  Stable plasmoids correspond to the lower branch.}
\end{figure*}

In Fig.~\ref{BLprop} we present the characteristics of axially
symmetric (snake-like BL~\cite{BycGolNik10}) and spherically
symmetric (spherical BL) plasmoids calculated on the basis of the
results of Sec.~\ref{NUM}.
The total energy of electric field inside a plasmoid (energy
density in 2D case) and the effective plasmoid radius are defined
as
\begin{equation}
  W_\mathrm{E} = \frac{1}{8\pi} \int E^2 \mathrm{d}V,
  \quad
  R_\mathrm{eff}^2 =
  \frac{1}{8 \pi W_\mathrm{E}} \int  r^2 E^2 \mathrm{d}V.
\end{equation}
The frequency of the total oscillation, including the main
harmonic oscillation which was separated in the derivation of
Eq.~\eqref{NLSE1}, is $f = f_p - |\Delta f|$, where $f_p = 2.8
\times \sqrt{n_5}\thinspace\text{MHz}$ is the Langmuir frequency.
Here we take into account that frequency shift $\Delta f$ for a
stable plasmoid is negative [see Figs.~\ref{2Dfig}(a)
and~\ref{3Dfig}(a)].

One can see in Fig.~\ref{BLprop} that in frames of our BL model we
predict the existence of low energy ($W_\mathrm{E} \gtrsim
\text{erg/cm}$ in 2D case and $W_\mathrm{E} \gtrsim
10^2\thinspace\text{erg}$ in 3D case) plasma structures with the
typical size $R_\mathrm{eff} \lesssim 1.5\thinspace\text{m}$. This
kind of plasmoids should appear in low density $n_e \sim
10^5\thinspace\text{cm}^{-3}$ and hot $T_e \sim
10^6\thinspace\text{K}$ plasma. Such a density of plasma can well
exist in the Earth ionosphere~\cite{Per97}. A linear lightning can
provide the plasma heating up to $T_e \sim 10^6\thinspace\text{K}$
since, e.g., $T_i \sim 10^4\thinspace\text{K}$ in the lightning
channel~\cite{RakUma06}. Thus we can assume that such high
temperatures of plasma can be present in some localized area of
ionosphere during a thunderstorm, making possible the existence of
plasma structures described in the present work. It is interesting
to notice that several meters objects were reported to appear in
the BL observations near airplanes~\cite{airplane}.

\begin{figure*}
  \centering
  \includegraphics[scale=1]{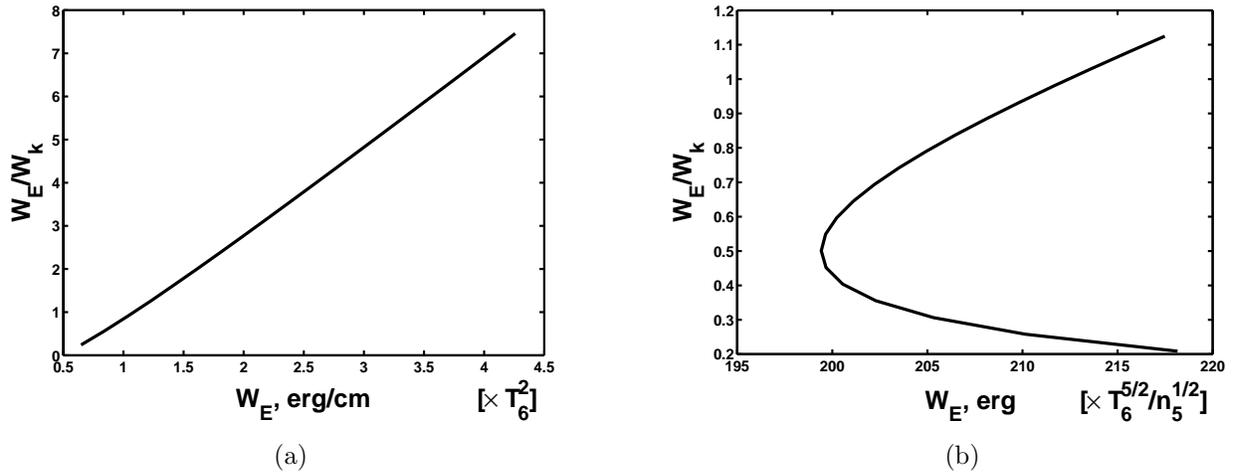}
  \caption{\label{stability}
  The ratio $W_\mathrm{E}/W_k$ vs. $W_\mathrm{E}$ for (a) 2D plasmoid and
  (b) 3D plasmoid. In the panel~(b) a stable plasmoid corresponds to
  the upper branch and an unstable one to the lower branch.}
\end{figure*}

To demonstrate the stability of the described plasmoids with
respect to decay, in Fig.~\ref{stability} we present the ratio
$W_\mathrm{E}/W_k$ versus $W_\mathrm{E}$, where $W_k = (3/2) n_0
T_e V_\mathrm{eff}$ is the total thermal energy of a plasmoid
(energy density in 2D case) and $V_\mathrm{eff}$ is the effective
volume of a plasmoid, which is equal to $\pi R_\mathrm{eff}^2$ for
$d=2$ and to $(4/3) \pi R_\mathrm{eff}^3$ for $d=3$.
Indeed, hot electrons with $T_e \sim 10^6\thinspace\text{K}$ could
just escape the plasmoid volume making it unstable. One can
however see in Fig.~\ref{stability} that for stable plasma
structures the ratio $W_\mathrm{E}/W_k$ has a tendency to increase
reaching the unit value at a certain soliton energy ($W_\mathrm{E}
\approx 1\thinspace\text{erg/cm}$ for $d=2$ and $W_\mathrm{E}
\approx 215\thinspace\text{erg}$ for $d=3$). It means that hot
electrons will not escape form the plasmoid volume as soon its
energy has this critical value. It should be noticed that an
unstable plasmoid, corresponding to the lower branch in
Fig.~\ref{stability}(b), will lose electrons since
$W_\mathrm{E}/W_k < 1$ always.

Comparing the predicted plasmoid properties with the
characteristics of circumterrestrial (not ionospheric) BL one can
say that our model is not directly applicable for the description
of such an object since its energy $\gtrsim 10\thinspace\text{kJ}$
and the size $\sim (20-50)\thinspace\text{cm}$ are beyond our
predictions. We can however use our model at the initial stages of
the plasmoid formation when the amplitude of the electric field is
so high since in Eq.~\eqref{NLSE1} we keep only cubic
nonlinearity. Higher nonlinear terms, which seem to be important
for denser plasma, can explain smaller size and bigger energy
content.

A natural plasmoid appearing in circumterrestrial atmosphere is
surrounded by the neutral gas. It means that plasma in the
interior of a plasmoid should be maintained in the state with
proper ionization during its lifetime. It is however
known~\cite{Smi93} that plasma of a low energy plasmoid, without
an internal energy source, will lose energy and recombine back to
a neutral gas in the millisecond time scale at the atmospheric
pressure. Thus the lifetime of such an object will be extremely
short. It was suggested that under certain conditions plasma can
reveal superconducting properties~\cite{supercond}. This mechanism
could prevent the energy losses and thus the recombination,
providing the long lifetime of a low energy plasmoid.

\section{Conclusion\label{CONCL}}

In conclusion we mention that in the present work we have studied
axially and spherically symmetric Langmiur solitons in warm
plasma. In Sec.~\ref{DYN} we started with the discussion of
radially symmetric oscillations of electrons in plasma and then
separated the motions on different time scales. In
Sec.~\ref{LOCNON} we briefly reviewed the previous works on the
theory of stable plasma structures which involved local electron
nonlinearities. Then, in Sec.~\ref{NONLOCNON} we have calculated
the contribution of the slowly varying electron density to the
electron pressure and derived, together with the results of
Ref.~\cite{DavYakZal05}, the new NLSE~\eqref{NLSE1} which accounts
for nonlocal electron nonlineariries. These additional nonlinear
terms do not disappear in one dimensional case. This fact was also
demonstrated in Appendix~\ref{LAGRANGE} using Lagrange variables.

The solutions of this new NLSE, rewritten in the equivalent
form~\eqref{NLSE3} for the radially symmetric case, have been
analyzed numerically in Sec.~\ref{NUM}. We presented the examples
of some of the solutions of Eq.~\eqref{NLSE3} [see
Fig.~\ref{2Dfig}(c,d) and Fig.~\ref{3Dfig}(c,d)] and analyzed
their stability. It has been found that solutions in 2D case are
stable, whereas in tree dimensions some unstable solitons can
exist. We also compared out results with Ref.~\cite{DavYakZal05},
where analogous NLSE was considered.

In Sec.~\ref{APPL} we suggested that the described axially and
spherically symmetric solitons can be realized in the form of a
natural plasmoid, a ball lightning. We have computed the
characteristics of such a plasma structure, like energy and radius
(see Fig.~\ref{BLprop}), predicted in frames of our model.
Comparing these characteristics with the properties of a natural
plasmoid one concludes that our model describes a low energy BL
which can exist in hot ionospheric plasma. Note that in many cases
BLs were observed form airplanes~\cite{airplane} at high
altitudes. We also examined the stability of plasmoids with
respect to decay because of the escape of hot electrons.

We have already considered a BL model based on radial oscillations
of electrons in our previous publications~\cite{DvoDvo,Dvo10}. In
the present work we performed more detailed analysis of electron
nonlinearities and showed that they play a crucial role for the
stability of a plasmoid. Although plasma structures described in
the present work does not reproduce all the properties of
circumterrestrial (not ionospheric)
BL~\cite{Ste99p11,BycGolNik10p214}, we can use our results for the
description of the BL formation, when the amplitude of the
electric field is not so big.

As we have found in Sec.~\ref{NUM} [see also
Fig.~\ref{BLprop}(a,c)], the total frequency of electron
oscillations is less than Langmuir frequency, since the frequency
shift $\Delta f$ is negative. It is important fact for the future
experimental studies of BL. For example, the electron density in a
linear lightning discharge can be $\sim
10^{17}\thinspace\text{cm}^{-3}$~\cite{RakUma06p163}, giving for
the plasma frequency a huge value of $\sim
10^3\thinspace\text{GHz}$. In a laboratory it is extremely
difficult to create a strong electromagnetic field of such a
frequency to generate BL. However, if the frequency of electron
oscillations has a tendency to decrease, it can facilitate the
plasmoid generation.


\begin{acknowledgments}
  This work has been supported by Conicyt (Chile), Programa
  Bicentenario PSD-91-2006. The author is thankful to
  S.~I.~Dvornikov,
  E.~A.~Kuznetsov, L.~Stodolsky, and A.~A.~Sukhorukov
  for helpful discussions and to
  Deutscher Akademischer Austausch Dienst for a grant.
\end{acknowledgments}

\appendix

\section{Lagrange variables\label{LAGRANGE}}

The propagation of waves in plasma was analyzed in
Sec.~\ref{LOCNON} using Euler variables which are more convenient
for the practical purposes. There is, however, a Lagrange approach
for the treatment of plasma waves. Instead of describing of plasma
characteristics, like velocity, density etc, in a certain point of
space, one can study the dependence of these quantities on the
initial coordinates of plasma particles.

Let us change the variables $(r,t) \to (\rho,\tau)$ in
Eqs.~\eqref{hydrodyneq} and~\eqref{rdependence} as
\begin{align}\label{Lagrvar}
  r = & \rho + \xi(\rho,\tau),
  \quad
  t = \tau,
  \notag
  \\
  \frac{\partial}{\partial \rho} = &
  \frac{\partial r}{\partial \rho}\frac{\partial}{\partial r},
  \quad
  \frac{\partial}{\partial \tau} =
  \frac{\partial}{\partial t} + v_e \frac{\partial}{\partial r},
\end{align}
where $\rho$ is the initial coordinate of an electron, $\xi$ is
deviation of an electron from the equilibrium, and $\tau$ is the
new temporal variable. An electron is supposed to be in the
equilibrium initially, $\xi(\rho,0) = 0$. In Eq.~\eqref{Lagrvar}
we use the definition of electron velocity, $\partial r / \partial
\tau =
\partial \xi / \partial \tau = v_e$. Unlike the Euler picture,
the continuity equation in Lagrange variables can be integrated
and it does not contain the time derivative,
\begin{equation}\label{Lagrcontinuity}
  n_0 \rho^{d-1} = n_e r^{d-1} \frac{\partial r}{\partial \rho}
\end{equation}
where $n_0 = n_e(\tau = 0)$ is the initial (unperturbed) electron
density, which is supposed to be uniform.

Using Eqs.~\eqref{Lagrvar} and~\eqref{Lagrcontinuity} we can
obtain from Eq.~\eqref{hydrodyneq} a single nonlinear equation for
electron velocity,
%
%
\begin{align}\label{Lagrve1}
  \ddot{v}_e & + \omega_p^2 \frac{n_i}{n_0} v_e + v_e \dot{v}_e \frac{d-1}{r}
  \notag
  \\
  & + 3 v_\mathrm{T}^2
  \bigg[
    \frac{2 r'' v_e'}{r'^3}-\frac{v_e''}{r'^2}+
    (d-1)
    \left(
      \frac{v_e}{r^2}-\frac{v_e'}{\rho r'^2}-\frac{v_e r''}{r r'^2}
    \right)
    \notag
    \\
    & +
    (d-1)^2 \frac{v_e}{r}
    \left(
      \frac{1}{\rho r'}-\frac{1}{r}
    \right)
  \bigg] = 0,
\end{align}
where a ``dot" and a ``prime" mean the derivatives with respect to
$\tau$ and $\rho$.
To derive Eq.~\eqref{Lagrve1} we assume that electrons obey the
adiabatic equation $p/n_e^{\gamma_e} = \text{const}$, where $p$ is
electron pressure [$p_{ij} = p \delta_{ij}$, see
Eq.~\eqref{pressure}] and $\gamma_e = 3$ since we study
electrostatic oscillations~\cite{Jac65}. Analogous assumptions
were taken in Ref.~\cite{InfRow87} to study the nonlinear
one-dimensional plasma waves in warm plasma within the Lagrange
picture.

Note that Eq.~\eqref{Lagrve1} is an exact one, which does not
suppose any expansion over a small parameter. We can, however,
discuss the small deviations from the equilibrium position,
%
%
\begin{align}\label{Lagrve2}
  \ddot{v}_e & + \omega_p^2
  \left(
    1 + \frac{n}{n_0}
  \right)  v_e -
  3 v_\mathrm{T}^2 \frac{\partial}{\partial \rho}
  \left[
    \frac{1}{\rho^{d-1}} \frac{\partial}{\partial \rho}
    \left(
      \rho^{d-1} v_e
    \right)
  \right]
  \notag
  \\
  & +
  v_e \dot{v}_e \frac{d-1}{\rho}
  +
  3 v_\mathrm{T}^2
  \bigg[
    2(\xi'' v_e' + v_e'' \xi')
    \notag
    \\
    & +
    (d-1)(d-3)\frac{v_e \xi}{\rho^3} +
    (d-1)
    \left(
      2\frac{v_e'\xi'}{\rho} - \frac{v_e \xi''}{\rho}
    \right)
    \notag
    \\
    & -
    (d-1)^2 \frac{v_e \xi'}{\rho^2}
  \bigg] = 0,
\end{align}
keeping only quadratic nonlinearities. As in Sec.~\ref{LOCNON},
$n= n_i - n_0$, stays for the perturbation of the ion density in
Eq.~\eqref{Lagrve2}.

It can be noticed that Eq.~\eqref{Lagrve2} has analogous structure
as Eq.~\eqref{NLSE1} before the separation of the main harmonic
(for the details see, e.g., Ref.~\cite{MusRubZak95}). We can see,
however, that nonlinear terms do not disappear completely at $d=1$
if we consider warm plasma with $T_e \neq 0$. One can also notice
that these nonvanishing nonlinear terms arise from the pressure
term in the initial plasma hydrodynamics
equations~\eqref{hydrodyneq}. Analogous conclusion was obtained in
Sec.~\ref{NONLOCNON} using Euler coordinates. Note that the
nonlinear plasma waves were also studied in
Refs.~\cite{InfRow87,Daw59}. It was found in Ref.~\cite{InfRow87}
that the nonlinear terms are important even for one dimensional
plasma oscillations for the case of nonzero electron temperature,
which is in agreement with our results [see Eqs.~\eqref{NLSE3}
and~\eqref{Lagrve2}].

\end{document}